\documentclass[lettersize,journal]{IEEEtran} 
\usepackage{booktabs}  
\usepackage{comment}
\usepackage[utf8]{inputenc} 
\usepackage[T1]{fontenc} 
\usepackage{geometry}
\usepackage{hyperref}
\usepackage{textcomp}
\usepackage{graphicx} 
\usepackage[font=scriptsize]{caption}
\usepackage{subcaption}
\usepackage{wrapfig} 
\usepackage{url}
\usepackage{hyperref}
\usepackage{indentfirst}
\usepackage{amsmath, amssymb,amsfonts}
\usepackage{xcolor}
\usepackage{color}
\usepackage{multicol}
\usepackage{cite}
\usepackage{array}
\usepackage{mathtools}
\usepackage{fancyhdr}
\usepackage{multicol}
\usepackage{nameref}
\usepackage{accents}
\usepackage{dsfont}
\usepackage{bbm}
\usepackage{algorithmic}
\usepackage{algorithm}
\usepackage{tikz}
\usetikzlibrary{decorations.pathreplacing}
\DeclarePairedDelimiterX{\inp}[2]{\langle}{\rangle}{#1, #2}
\allowdisplaybreaks[3]
\DeclareMathAlphabet{\mymathbb}{U}{BOONDOX-ds}{m}{n}

\newcommand{\utwi}[1]{\mbox{\boldmath $#1$}}

\newcommand{\diag}{\mathsf{diag}}

\renewcommand{\tilde}{\widetilde}

\newcommand{\cN}{{\cal N}}

\newcommand{\cA}{{\cal A}}

\newcommand{\cT}{{\cal T}}

\newcommand{\cV}{{\cal V}}
\newcommand{\cX}{{\cal X}}

\newcommand{\cW}{{\cal W}}
\newcommand{\cK}{{\cal K}}

\newcommand{\bp}{{\bf p}}
\newcommand{\bq}{{\bf q}}

\newcommand{\bs}{{\bf s}}
\newcommand{\bx}{{\bf x}}
\newcommand{\bu}{{\bf u}}
\newcommand{\bv}{{\bf v}}

\newcommand{\bz}{{\bf z}}
\newcommand{\by}{{\bf y}}

\newcommand{\bV}{{\bf V}}

\newcommand{\bdelta}{{\utwi{\delta}}}

\newcommand{\blambda}{{\utwi{\lambda}}}

\newcommand{\bnu}{{\utwi{\nu}}}

\newcommand{\brho}{{\utwi{\rho}}}

\newcommand{\one}{\mathds{1}}

\newcommand{\real}{\mathbb{R}}

\DeclarePairedDelimiterX{\norm}[1]{\lVert}{\rVert}{#1}

\newtheorem{proposition}{Proposition}

\newtheorem{problem}{Problem}

\newtheorem{assumption}{Assumption}
\newtheorem{remark}{Remark}

\title{Incremental Volt/Var Control for Distribution Networks via Chance-Constrained Optimization}

\author{Antonin Colot$^*$, Elisabetta Perotti$^*$, Mevludin Glavic, and Emiliano Dall'Anese 
\thanks{$^*$ Equal contribution of the authors.}
\thanks{The work of A. Colot was supported by the Research Fellow Fellowship of the F.R.S-FNRS. The work of E. Dall'Anese was supported in part by the National Science Foundation (NSF) award 2448264. The work of E. Perotti was supported by Schmidt Science Fellows, in partnership with the Rhodes Trust. }
\thanks{A. Colot and M. Glavic are with the Montefiore Institute, University of Li\`{e}ge (email: antonin.colot@uliege.be; mevludin.glavic@uliege.be).}
\thanks{E. Perotti is with the Department of ECEE, University of Colorado Boulder (email: elisabetta.perotti@colorado.edu). }
\thanks{E. Dall'Anese is with the Department of Electrical and Computer
Engineering and the Division of Systems Engineering, Boston University (email: edallane@bu.edu).}
}
\date{\today}

\begin{document}
\maketitle
\begin{abstract}
This paper considers an incremental Volt/Var control scheme for distribution systems with high integration of inverter-interfaced distributed generation (such as photovoltaic systems). The incremental Volt/Var controller is implemented with the objective of minimizing reactive power usage while maintaining voltages within safe limits sufficiently often. To this end, the parameters of the incremental Volt/Var controller are obtained by solving a chance-constrained optimization problem, where constraints are designed to ensure that voltage violations do not occur more often than a pre-specified probability. This approach leads to cost savings in a controlled, predictable way, while still avoiding significant over- or under-voltage issues. The proposed chance-constrained problem is solved using a successive convex approximation method. Once the gains are broadcast to the inverters, no additional communication is required since the controller is implemented locally at the inverters. The proposed method is successfully tested on a low-voltage single-phase 42-nodes network and on the three-phase unbalanced IEEE 123-node test system.
\end{abstract}


\section{Introduction}
The increasing integration of distributed energy resources (DERs) is driving a paradigm shift in electrical power networks, moving away from centralized power plants and embracing decentralized energy systems.
The increasing integration of inverter-based renewable energy resources (RESs) into the electricity generation mix, driven by the need to meet climate targets, presents significant challenges across all aspects of electrical distribution networks (DNs), from long-term planning to real-time control~\cite{walling2008summary,alam2020high}. 
In this paper, we seek new solution approaches to the voltage regulation problem in DNs. Traditionally, voltage regulation in DNs was achieved using load tap changers or switchgears. However, the increasing
variability introduced by DERs can shorten their lifespan, and they may become insufficient to regulate voltages~\cite{mcdonald2013solar}. 
On the other hand, continuous improvements in power electronics converters create new possibilities for the control of  DERs~\cite{antoniadou2017distributed}, thus enabling new methods for real-time voltage regulation.

Leveraging the inverters' capabilities, we propose a two-level Volt/Var control strategy, driven by the following practical  and architectural considerations:

\noindent \emph{(a)} Piecewise linear droop Volt/Var rules, as suggested by the IEEE 1547 standard, offer fast response to deviations of local voltage magnitude~\cite{Zhu2016,Zhou2021}.

\noindent \emph{(b)} Local incremental Volt/Var methods offer advantages over the piecewise linear droop Volt/Var rule in terms of utilization of reactive power resources and avoidance of possible oscillations~\cite{Jahangiri2013,Farivar2015}. 

\noindent \emph{(c)} Local Volt/Var methods often do not guarantee voltage regulation; communication and coordination are typically required~\cite{Bolognani2019}.

The method proposed in this paper enjoys the advantages of the local incremental controllers (acting at a fast time scale locally), with regulation performance ensured by designing the control parameters via an optimization-based task (solved occasionally at the higher level). In particular, the optimization-based task allows us to specify the maximum amount of voltage violation that is tolerated. The only communication needed is between a central aggregator where the controller parameters are computed and the inverters, which receive the updated controller parameters from the aggregator. In particular, we do not require \textit{real-time} communication as the controller parameters are obtained using forecasts for which uncertainties are taken into account in the optimization-based task.

\subsection{Literature Review}
Volt/Var controllers determine reactive power injections from, e.g., a static function of the local voltage measurements. The slopes of individual Volt/Var curves can be tuned to achieve various objectives by solving appropriate optimization problems~\cite{Baker2017}. Since these static feedback laws can lead to oscillatory behaviors~\cite{Jahangiri2013}, incremental strategies, based on voltage measurements and the past reactive power setpoint~\cite{Farivar2015}, are generally favored.   
Chance-constrained approaches to design optimal rules for \textit{non}-incremental Volt/Var control are proposed in~\cite{Wei2023},~\cite{nazir2019}. These works consider a separate set of gains for each DER. This may be challenging in practice, as it requires knowledge of each DER location and an advanced communication infrastructure to properly dispatch the gains.

\begin{figure*}[t!] 
\centering
\begin{subfigure}[t]{0.5\textwidth} \centering
  \includegraphics[width=\textwidth]{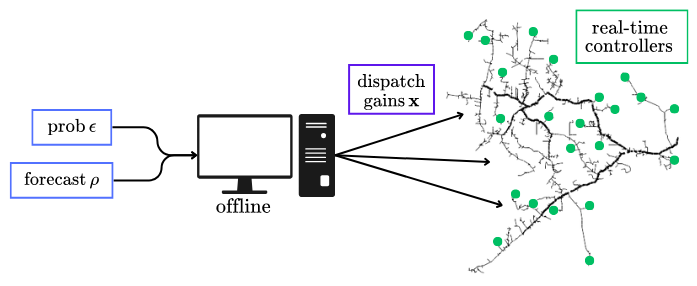}
 \caption{}
\end{subfigure}
\hspace{1.5em}
\begin{subfigure}[t]{0.2\textwidth} \centering
 \includegraphics[width=\textwidth]{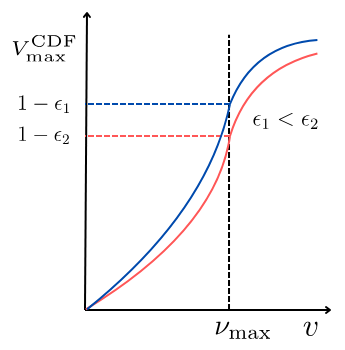}
 \caption{} 
\end{subfigure}
\caption{(a) Proposed voltage regulation strategy. Based on the forecast $\brho$ and the probability $\epsilon$ to violate voltage limits, the controller gains $\bx$ are computed centrally and then broadcasted to the controllable DERs. (b) Illustrative explanation of the impact of the parameter $\epsilon$ on the total amount of voltage violations. A smaller $\epsilon$ results in a more constrained optimization problem and therefore in fewer voltage violations.}
\label{fig:cartoons}
\vspace{-.2cm}
\end{figure*}

We also mention some representative works in the context of data-driven methods to learn a Volt/Var controller~\cite{Karagiannopoulos2019}. For learning-based strategies, it is often difficult to analyze the closed-loop stability~\cite{eggli2020stability}. Some exceptions are, e.g.~\cite{Gupta2023,gupta2024optimal}; however, the controllers in these works do not minimize reactive power usage, potentially causing additional losses in the DN. In~\cite{yuan2023learning}, closed-loop stability for a general class of Volt/Var controllers is guaranteed but historical data are needed for training the learning-based controllers.
In the context of learning-based Volt/Var, however, it is difficult to account for network topology changes. Indeed, changes in topology may require collecting new data and re-training learning-based controllers, which is time-consuming. In~\cite{SERNATORRE2022108575}, authors propose a linear Volt/Var controller and a methodology for adapting the controller to varying network topologies. However, this requires solving multiple instances of the optimal power flow problem.

\subsection{Statement of Contributions} 
The proposed approach is illustrated in Fig.~\ref{fig:cartoons}. This approach can be viewed as a two-level control where: (1)~the coefficients and gains of the local controllers are computed centrally (e.g., in an Advanced Distribution Management System) and then broadcasted to the local controllers (the coefficients and gains are updated at a slow time scale, for example at hourly intervals); and (2) the DERs implement the controllers, with the received coefficients and gains, for real-time voltage regulation. In real-time, the control is fully decentralized, in the sense that the controller is implemented locally at each DER without DER-to-DER communications. The gains are computed by solving an optimization problem that embeds specifications on the maximum amount of voltage violation that is tolerated.

The contributions of this paper are as follows:
\begin{itemize}
    \item A chance-constrained optimization problem is formulated to ensure that voltage violations cannot occur more often than according to a given probability. 
    \item A single set of gains suitable for all DERs is determined, thus simplyfing the optimization problem to be solved at the upper level and keeping the communication infrastructure requirements low. Furthermore, \textit{fairness} in sharing control efforts is implicitly embedded by having the same gains for all DERs.
    \item With respect to a standard Volt/Var controller, this approach aims to minimize the generator’s reactive power injections. The controller gains are determined in advance based on forecast data for power generation and loads. To account for a variety of possible distributions of the forecasting errors, a conservative approximation of the chance constraints is derived~\cite{Nemirovski2007}.
    \item There is no need of historical data of any type. Therefore, not only can this approach easily handle planned topology changes but, even in case of unplanned changes, it is still reliable since network conditions are taken into account as feedback. 
    \item  This approach uses an algorithm based on Successive Convex Approximation (SCA) methods~\cite{Scutari2014} to derive a convex, and therefore tractable formulation of our chance-constrained problem.
\end{itemize} 
To the authors’ knowledge, this is the first work that relies on chance-constrained optimization to determine the gains of an incremental Volt/Var controller.

\subsection{Outline}
In Section \ref{sec:prob_formulation}, we define the DN model and formulate the problem. Section \ref{sec:feedback_contr_design} introduces our incremental local Volt/Var controller and analyses its stability. Section \ref{sec:gains_design} presents a strategy to determine the controller gains that optimize the performance of our controller. Section \ref{sec:implementation_contr} provides details regarding the implementation of the feedback controllers.
Numerical simulations in Section \ref{sec:num_exp} compare the proposed approach to benchmark methods. Section \ref{sec:conclusion}
concludes the paper.

\section{Problem formulation}\label{sec:prob_formulation}
\subsection{Power system model}\label{subsec:power_system_model}
Consider a balanced three-phase power distribution network\footnote{\emph{Notation. Upper-case (lower-case boldface) letters are used for matrices (column vectors); $(.)^\top$ denotes the transposition, $(.)^*$ denotes the complex conjugate and $(.)^{-1}$ denotes the inverse matrix; 
$j$ the imaginary unit and $|.|$ the absolute value of a number. If we consider a given vector $\bx \in \real^{N}$, $\diag(\cdot)$ returns a $N\times N$ matrix with the element of $\bx$ in its diagonal. For vectors \( \bx \in \mathbb{R}^n \) and \( \bu \in \mathbb{R}^m \),
\( \| \bx \| \) denotes the $\ell_2$-norm and  \( (\bx,\bu) \in \mathbb{R}^{n + m}\) denotes their vector concatenation.
For a matrix $A \in \mathbb{R}^{N\times N}$, $\|A\|$ denotes the induced $\ell_2$-norm.
We denote as $\mathbf{0}$ a vector with all zeros (the dimensions will be clear from the context). We denote $\mathbb{R}_{> 0},\mathbb{R}_{\geq 0}$ and $\mathbb{R}_{\leq 0}$ the set of strictly positive, positive and negative real numbers. $\mathds{1}$ denotes the $N\times1$ column vector and $\mathds{I}$ the $N\times N$ identity matrix.}} with $N+1$ nodes and hosting $G$ DERs. The node 0 is taken to be the point of common coupling, while $\mathcal{N} := \{1,...,N\}$ is the set of remaining nodes. We consider a phasor model of the single-phase equivalent DN where we define $u_k = v_k e^{j\delta_k} \in \mathbb{C}$ the voltage at node $k$. The voltage at node 0 is set to $u_0 = V_0$. Using Ohm's and Kirchhoff's laws, one has the usual relationship:
\begin{equation}\label{eq:pf_eq}
    \bs = \diag{(\bu)}(\by^*V_0+Y^*\bu^*),
\end{equation}
where $\bu := \{u_k\}_{k\in \cN}$ collect the voltages at every node, and $\bs = \bp_{\mathrm{av}}-\bp_{\mathrm{l}} + j(\cA \tilde\bq-\bq_{\mathrm{l}})\in \mathbb{C}^N$ is the net power injection at nodes $n \in \cN$. The network bus admittance matrix is partitioned such that $\by \in \mathbb{C}^{N}, Y \in \mathbb{C}^{N\times N}$. We define $\bp_{\mathrm{av}},\bp_{\mathrm{l}},\bq_{\mathrm{l}} \in \mathbb{R}^{N}$ as the vectors collecting the non-controllable active power injection, the active and reactive power consumption at nodes $n\in \cN$, respectively.  We define $\tilde\bq \in \mathbb{R}^G$ as the vector collecting the controllable reactive power of the $G$ DERs. The matrix $\cA \in \mathbb{R}^{N\times G}$ maps the index of a DER to the node where it is located. 

Equation~\eqref{eq:pf_eq} describes the power flow equations. For a given vector of net power injection $\bs$, one can solve this non-linear system of equations using numerical methods to find the vector of voltage phasors $\bu$. Notice that the system of equations~\eqref{eq:pf_eq} may have zero, one or many solutions. For the rest of this paper, we make the following assumption.

\begin{assumption}[\textit{Existence and uniqueness of a practical solution of the power flow equations}]\label{assumption:pf_eq}
There exists at least one solution to the power flow equations~\eqref{eq:pf_eq}. If multiple solutions exist, we only consider the \emph{practical} solution, i.e., in the neighborhood of the nominal voltage profile, we pick the high voltage, and small line currents solution. \hfill $\Box$
\end{assumption}

Let $\bz := (\bp_{\text{av}},\bp_{\text{l}},\bq_{\text{l}})$ be the concatenation of the non-controllable powers, the algebraic map $H := \mathbb{R}^{G+3 N} \rightarrow \mathbb{R}^{N}$ and $\bv := \{v_k\}_{k\in \cN}$ the vector collecting the voltage magnitudes at every node. For convenience, we denote $\bv = H(\tilde\bq,\bz)$ where $H$ relates the net power injections to the \emph{practical} solution of the power flow equations defined in~\eqref{eq:pf_eq}. Although the analytical form of the map $H$ is not known, its existence and uniqueness have been discussed for balanced~\cite{bolognani2015existence,wang2016explicit} and multi-phase DNs~\cite{wang2017existence}. 

\subsection{Problem setup}\label{sec: problem_setup}

The deployment of inverter-interfaced generation in DNs might induce voltage quality issues. However, DERs can also be used to provide voltage regulation services via, e.g., reactive power compensation. Ideally, one wants to minimize the total reactive power usage while maintaining voltages inside a given feasible set. We formulate an optimal reactive power flow problem as follows: 

\begin{equation}\label{eq:P0}
    \begin{aligned}
        (\text{P0}(t)) \quad \min_{\tilde\bq \in \mathcal{Q}}  & \quad f(\tilde\bq) \\
        \textrm{s.t.} & \quad H(\tilde\bq,\bz(t)) \in \mathcal{V},
    \end{aligned}
\end{equation}
where $\mathcal{Q} \subset \mathbb{R}^{G}$ is the set of feasible values for the reactive powers, $\mathcal{V} \subset \mathbb{R}^{N}$ is the set of feasible values for voltage magnitudes, and $f:\mathbb{R}^{G}\rightarrow\mathbb{R}$ is a differentiable cost function. Notice that (P0($t$)) is a time-varying optimization problem. Indeed, the non-controllable power injections $\bz(t)$ vary with time as they depend on users' habits and weather conditions. Therefore, also the optimal reactive power injections are time dependent; their time-scale is determined by the variability of the non-controllable power injections and is usually within seconds~\cite{dall2016optimal}.  
However, collecting data, solving problem (P0($t$)) and broadcasting the setpoints to the inverters every second is challenging because of the non-linear nature of the power flow equations~\eqref{eq:pf_eq} represented by $H$, and the communication burden associated with large DNs. This paper aims to solve the following problem.
\begin{problem}\label{prob: main}
Design feedback controllers to approximate the solution of (P0($t$)) with limited computational resources and in a decentralized fashion, i.e., each controller uses only local voltage measurements to compute its reactive power output and the operation does not necessitate continuous communications with neighbors or a centralized entity. \hfill $\Box$
\end{problem}

To outline the proposed framework, we consider the case where there is one DER per node in the network and the map $H$ does not change over time. Additionally, we make the following assumption.
\begin{assumption}[Feasibility]\label{assumption: main}
For any $\bq \in \mathcal{Q}$, there exists a solution such that $\bv(t) \in \mathcal{V}$. \hfill $\Box$   
\end{assumption}

This assumption ensures that there is enough reactive power reserve to regulate the voltages. We will review this assumption and our setup choices later, and discuss how they could be relaxed.

\section{Design of feedback controllers}\label{sec:feedback_contr_design}
Within the next section, we identify three different time scales. In order from the shortest to the longest, they correspond to the time scales that characterize the following events: i) controller law updates, ii) forecast updates and iii) controller gains updates.

\paragraph{Incremental Volt/Var control}\label{subsec:inc_volt_var} 
To begin with, we consider only the shortest time scale, which is related to the controller law updates.
Let us discretize the temporal domain as $t = k\tau$, where $k \in \mathbb{N}_+$ and $\tau \in \mathbb{R}_{> 0}$ is a given time interval, small enough to resolve variations in the time-dependent disturbances, i.e., less than a second. We consider the following feedback controller:
\begin{equation}
        \bq_{k+1}=\bq_k+\eta(\mathds{1}-\bnu_k)-(1-\eta)\alpha\bq_k,\label{eq:controller} 
\end{equation}
where $\alpha \in \mathbb{R}_{\geq 0}$, $\eta \in [0,1]$, and  $\bnu_k = X\bq_k + \brho_k$ with $X\in \mathbb{R}^{N\times N}$ is a linear approximation of the power flow equations and $\brho_k = H(\mathbf{0},\bz_k)$ denotes the voltage profile obtained by setting the controllable reactive powers to 0.
The linearized power flow equations can be derived from the bus injection model \cite{dall2016optimal}, or from the branch flow model \cite{LiQuDahleh2014}. For the rest of this paper, we consider the linearized power flow equations based on the branch flow model~\cite{baran1989network}, since it guarantees $X$ being positive definite~\cite{farivar2013equilibrium}. This approximation has been used in~\cite{LiQuDahleh2014,eggli2020stability}, and its quality has been discussed in~\cite{farivar2013equilibrium}. Notice that $\brho_k = H(\mathbf{0},\bz_k)$ is derived from the true power flow equations. 
Substituting this approximation in~\eqref{eq:controller}, the controller  can be written as:
\begin{equation}\label{eq:controller_la}
        \bq_{k+1} =A(\eta,\alpha)\bq_k+B(\eta,\brho),
\end{equation}
where $A(\eta,\alpha) := (1 -(1-\eta)\alpha)\mathds{I} -\eta X$ and $B(\eta,\brho)  := \eta (\mathds{1}-\brho)$. 

\paragraph{Existence and uniqueness of the equilibrium}

Denoting $\brho = \brho_k$ for a given $k$, the equilibrium for~\eqref{eq:controller_la} is defined as:
\begin{equation}\label{eq:equilibrium}
    \begin{aligned}
        \bq^*  &=\left[\eta X + (1-\eta)\alpha\mathds{I}\right]^{-1}\eta(\mathds{1}-\brho)\\
        \bnu^* &= X\left[\eta X + (1-\eta)\alpha\mathds{I}\right]^{-1}\eta(\mathds{1}-\brho) + \brho.
    \end{aligned}
\end{equation}
Since $X$ is positive definite and $\alpha \in \mathbb{R}_{\geq 0}$, $\eta \in [0,1]$, the matrix $\eta X + (1-\eta)\alpha\mathds{I}$ in~\eqref{eq:equilibrium} is always invertible and the equilibrium is unique.
One can check that for $\eta = 0$ and $\alpha > 0$, $\bq^*  =\mathbf{0}$ and $\bnu^* = \brho$, while for $\eta > 0$ and $\alpha = 0$ or $\eta=1$, $\bq^*  =X^{-1}(\mathds{1}-\brho)$ and $\bnu^* = \mathds{1}$.

Increasing the gain $\alpha$ decreases the use of reactive power, while increasing the gain $\eta$ steers the voltage magnitudes to the nominal voltage profile. 

\paragraph{Stability analysis}
 The controller~\eqref{eq:controller_la} is asymptotically stable if and only if $\rho(A(\eta,\alpha))<1$, where $\rho(\cdot)$ denotes the spectral radius.
This condition is verified if:
\begin{equation}\label{eq:P1_stab_cstr_full}
    \mathbf{0} < (1-\eta)\alpha\mathds{1} + \eta \blambda_X < 2\mathds{1},
\end{equation}
where $\blambda_X \in \mathbb{R}^{N}$ is the vector containing the eigenvalues of the matrix $X$.
Moreover the matrix $X$ is positive definite by construction, and since $\eta, \alpha$ satisfy $\alpha \in \mathbb{R}_{\geq 0}$, $\eta \in [0,1]$, we always have $(1-\eta)\alpha\mathds{1} + \eta \blambda_X \geq \mathbf{0}$. The equality holds only if $\eta = \alpha = 0$, which guarantees a stable controller since $\bq_{k+1} = \bq_{k}$.

\paragraph{Multi-phase unbalanced distribution networks}\label{subsec:multi_phase}
DNs are often highly unbalanced, and may have different connection configurations (wye-connected or delta-connected). In order to study the stability properties of the controller~\eqref{eq:controller} for unbalanced DNs, one would need to use a different approximation of the power flow equations, e.g.,~\cite{wang2017existence,kekatos2015voltage}. In our formulation, we leverage the positive definite property of the matrix $X$ to show the uniqueness and existence of the equilibrium, as well as the stability of the controller. To adapt this framework to unbalanced DNs, one would need to study the properties of the new matrix $X$ to characterize the equilibrium and stability properties of the controller~\eqref{eq:controller}. However, as mentioned in~\cite{kekatos2015voltage}, due to the structure of the distribution lines, $X$ is often positive definite for unbalanced networks. In such cases, our methodology can be readily applied to multi-phase networks.

\section{Design of the controller gains}\label{sec:gains_design}
The performance of the controller~\eqref{eq:controller_la} depends on the choice of $\eta$ and $\alpha$.  In the following section, we introduce an optimization-based method to design the gains.
\subsection{Time-varying formulation}
Given a matrix $X$, a time-varying vector $\brho_k = H(\mathbf{0},\bz_k)$, and feasible sets $\mathcal{Q}$ and $\mathcal{V}$, we formulate the following problem at time $k \tau$: 
\begin{subequations}\label{eq:P1}
    \begin{align}
        (\text{P1}_k) \quad  \min_{\alpha \in \mathbb{R}_{\geq 0},  \eta \in [0,1]} & \quad \|\bq_k(\alpha,\eta)\|^2\\
        \textrm{s.t.} & \quad \bq_k(\alpha,\eta) \in \mathcal{Q}\\
                      & \quad X\bq_k(\alpha,\eta)+\brho_k \in \cV\\
                      & \quad (1-\eta)\alpha\mathds{1} + \eta \blambda_X < 2\mathds{1} \label{eq:P1_stab_cstr}
    \end{align}
\end{subequations}
where $\bq_k(\alpha,\eta) = [\eta X + (1-\eta)\alpha\mathbb{I}]^{-1}\eta(\mathds{1}-\brho_k)$ is a non-linear function of $\eta$ and $\alpha$. For a given time $k\tau$, the goal is to minimize the reactive power usage $ \|\bq_k(\alpha,\eta)\|^2$, while satisfying operational constraints, by appropriately selecting the gains $\eta$ and $\alpha$. 

An alternative formulation of problem (P1$_k$) can be derived using an appropriate change of variable. This facilitates the application of the successive convex approximation strategy presented in Section~\ref{sec:convexification}. Therefore, we introduce the optimization variable $\bx = [\frac{\alpha}{\eta},-\alpha]^\top \in \mathbb{R}_{\geq 0} \times \mathbb{R}_{\leq 0}$, rewrite $\bq_k(\bx) = \left[X + \mathbf{1}^\top \bx \mathds{I}\right]^{-1}(\mathds{1}-\brho_k)$, specify $\mathcal{Q}$ and $\cV$ in terms of box constraints, and reformulate the problem ($\text{P1}_k$) as
\begin{equation}\label{eq:P2}
    \begin{aligned}
        (\text{P2}_k) \quad \min_{\bx} & \quad h_0^k(\bx) \\
        \textrm{s.t.} 
        & \quad h_i^k(\bx) \leq 0\quad \forall i \in \{1,...,8\}\\
    \end{aligned}
\end{equation}
where 
 \begin{equation}\label{eq:P2_h_def}
    \begin{aligned}
        &h_0^k(\bx) = \|\bq_k(\bx)\|^2,\quad h_1^k(\bx) = \bq_k(\bx) - \bq_{\mathrm{max}},\\
        &h_2^k(\bx) = -\bq_k(\bx) + \bq_{\mathrm{min}},\\
        &h_3^k(\bx) = X\bq_k(\bx) + \brho_k - \bnu_{\mathrm{max}},\\
        &h_4^k(\bx) = -X\bq_k(\bx) - \brho_k + \bnu_{\mathrm{min}},\\
        &h_5(\bx) = \left(\mathbf{1}^\top \bx - 2\right) \mathds{1} + \blambda_X,\\
        &h_6(\bx) = - \mathbf{1}^\top\bx,\quad h_7(\bx) = -x_1,\quad h_8(\bx) = x_2,
    \end{aligned}
\end{equation}
 with $x_1, \,x_2$ scalar components of $\bx$ and $\mathbf{1} = [1,1]^\top \in \mathbb{R}^2$. The objective function $h_0$ is a scalar function, as well as $h_6,\,h_7$ and $h_8$. All other constraint functions $h_i$ with $i=1,...,5$ are vector-valued. Another advantage of performing the change of variables is that function $h_5(\bx)$ gives a tighter bound on \eqref{eq:P1_stab_cstr} with respect to the old formulation, with the equality reached for $\eta=1$. The controller gains derived from $(\text{P2}_k)$ ensure asymptotic stability of the controller defined in \eqref{eq:controller} as long as $\eta<1$. The conditions $\eta<1$ is always verified as $\eta=1$ implies $\bnu^*=\one$ which leads to a suboptimal solution (unless $\bnu_{\mathrm{min}} = \bnu_{\mathrm{max}} = \one$).

As explained in Section~\ref{sec: problem_setup},  collecting measurements at every nodes, solving $(\text{P2}_k)$ and then dispatching the controller gains to the controllable DERs in \textit{real-time} is unfeasible because of the communication burden and the computational time required to solve $(\text{P2}_k)$.
One could envision solving $(\text{P2}_k)$ for every time $k\tau$ using forecasts of $\brho_k$. However, it is not realistic to have such frequent forecast updates, since $\tau$ should be sufficiently small to cope with the DN dynamics. Furthermore, $\brho_k$ is affected by large uncertainties as it inherits them from $\bz_k$ through $\brho_k = H(\mathbf{0},\bz_k)$. Moreover, we would like to find optimal controller gains $\eta$ and $\alpha$ over a longer time period, to avoid broadcasting new values at every time $k\tau$, or to avoid storing a large number of gains in each controller. 
In the next section we address this issue by reformulating our problem in a chance-constrained fashion. 

\begin{remark}{\emph{(Choice of the coefficients)}}
In principle, we could use a grid search approach for $\eta$ and $\alpha$ in order to find the optimal pair of values that satisfies the constraints and
minimize the cost. However, we choose to take an optimization-based approach since it is not straightforward how to select the granularity to perform the screening over the values of $\alpha$ and $\eta$. Moreover, while $\eta$ is in the closed interval $[0,1]$, the only requirement on the coefficient $\alpha$ is to be  non-negative. Therefore, one key question pertains to how to select an interval $[0, \bar{\alpha}]$ for $\alpha$ when building a grid for $[0,1] \times [0, \bar{\alpha}]$. Setting $\bar{\alpha}$ requires an apriori knowledge of the locations of the minima of the function to be minimized, which we assume is not available. \hfill $\Box$
\end{remark}

\begin{figure}
    \centering
    \scalebox{0.8}{
    \begin{tikzpicture}
\draw[thick,->] (0,0) -- (9.5,0) node[anchor=north west] {$k\tau$};

    \foreach \x in {0,1,2,3,4,5,6,7,8,9}
   \draw (\x cm,1pt) -- (\x cm,-1pt) node[anchor=north] {$\x$};
\draw [decorate,decoration={brace,amplitude=5pt,mirror,raise=4ex}]
  (0,0) -- (3,0) node[midway,yshift=-3em]{$\Delta\tau$};
  \draw [decorate,decoration={brace,amplitude=5pt,mirror,raise=4ex}]
  (3,0) -- (6,0) node[midway,yshift=-3em]{$\Delta\tau$};
  \draw [decorate,decoration={brace,amplitude=5pt,mirror,raise=4ex}]
  (6,0) -- (9,0) node[midway,yshift=-3em]{$\Delta\tau$};
  \draw [decorate,decoration={brace,amplitude=5pt,raise=3ex}]
  (0,0) -- (9,0) node[midway,yshift=3em]{$T\tau$};
\end{tikzpicture}}
    \caption{Comparison between different time scales of the problem, assuming $\tau = 1s$.}
    \label{fig:timeScales}
\end{figure} 

\subsection{Chance-constrained formulation}
Given that our forecast $\brho_k = H(\mathbf{0},\bz_k)$ is subject to uncertainty 
in the vector $\bz_k$, we implement probabilistic constraints in our optimization problem in order to enforce voltage regulation with prescribed probability. 
The available DERs' powers at time $k\tau$ are modeled by $\bp_{\text{av},k}=\Bar{\bp}_{\text{av},k}+\bdelta_{\text{av},k}$, while the active and reactive
loads are expressed as $\bp_{\text{l},k}=\Bar{\bp}_{\text{l},k}+\bdelta_{\text{pl},k}$ and $\bq_{\text{l},k}=\Bar{\bq}_{\text{l},k}+\bdelta_{\text{ql},k}$, respectively. Writing $\bdelta_{k} := (\bdelta_{\text{av},k},\bdelta_{\text{pl},k},\bdelta_{\text{ql},k})$, and $\Bar \bz_k := (\Bar{\bp}_{\text{av},k},\Bar{\bp}_{\text{l},k},\Bar{\bq}_{\text{l},k})$, we have $\brho_k = H(\mathbf{0},\Bar \bz_k+\bdelta_{k})$ where $\bdelta_{k}$ follows a given distribution function. 

It is reasonable to assume that a new forecast for $\brho_k$ will be available after a certain time interval $\Delta\tau$ with $\Delta \in \mathbb{N}_+$ (for example, $\Delta$ could be such that $\Delta\tau = 15$ minutes). Concretely, this means that within a time window of duration $\Delta\tau$, $\brho_k$ will be the same, regardless of the value of the index $k$. Therefore, we will drop the index $k$ in the following, and consider $\brho^m$ instead to underline that the forecast $\brho^m$ will be updated at each time $t=m\Delta\tau$ with $m\in\mathbb{N}_+$. This introduces a longer time scale, whose magnitude is related to how often forecast updates occur.

Ultimately, our goal is to determine controller gains to be deployed over an even longer time interval  $T \tau$  where $T = b \Delta$ with $b \in \mathbb{N}_+$, e.g., $b$ is such that $T\tau = 1$ hour. Fig.~\ref{fig:timeScales} provides a visual representation of the relationship between the three different time scales relevant to this problem.

We consider an extension of $(\text{P2}_k)$ as a multi-period optimization problem: 
\begin{equation}\label{eq:P3}
    \begin{aligned}
        (\text{P3}) \quad  \min_{\bx} & \quad \sum_{m=1}^{b}\mathbb{E}\{h_0(\bx;\brho^m)\} \\
        \textrm{s.t.}   & \quad \mathrm{Pr}\{h_{i,n}(\bx;\brho^m)\leq 0\} \geq 1-\epsilon_i\\
 &\qquad \forall i \in \{1,2,3,4\}, n \in \cN,\, m \in \{1,..,b\} \\
                         &\quad h_{i}(\bx) \leq 0 \quad \forall i \in \{5,6,7,8\}\\
    \end{aligned}
\end{equation}
where we sum over $b$ intervals of magnitude $\Delta$, corresponding in total to a period of time $T\tau$. Notice that for $b = 1$, we recover a single-interval formulation. $\mathrm{Pr}\{A\}$ denotes probability of an event $A$ to happen, meaning that in problem (P3) the constraints $h_{i}(\bx;\brho^m)\le0$ for $i=1,2,3,4$ are satisfied with a probability $1-\epsilon_i$, where $\epsilon_i \in (0,1)$. Solving problem (P3) for the optimization variable $\bx$, we can retrieve the values of the gains $\alpha$ and $\eta$ to be deployed during a time interval of length $T\tau$. When following this approach we lose optimality in exchange for convenience: (P3) can be solved using a coarser forecast and controller gains are designed to cover wider time windows, which is a great advantage from a practical point of view.
Finally, notice that the objective function $h_0(\bx;\brho^m)$ also depends on $\brho^m$, thus we minimize its expected value.

We seek a tractable  approximation for the chance constraints in \eqref{eq:P3} since we do not know the probability distribution function of $\bdelta_k$, neither the map $H$. The chance constraints to be approximated are of the form $\mathrm{Pr}\{h(\bx;\brho)\le 0\} \ge 1-\epsilon$, where the function $h(\bx,\brho)$ depends on the optimization variable $\bx$ and the random vector $\brho$.
Consider $\psi(x)=[1+x]_+$, where $[x]_+:=\max\{x,0\}$, a so-called generating function $\psi: \mathbb{R} \rightarrow \mathbb{R}$ nonnegative, non-decreasing, and convex that satisfies the conditions $\psi(x)>\psi(0) \, \forall x>0$ and $\psi(0)=1$. Given a positive scalar $z>0$, we have that the following bound holds for all $z>0$ and $\bx$ \cite{Nemirovski2007}:
\begin{equation}\label{eq:approx_max}
\inf_{z\in\mathbb{R}}\{\mathbb{E}_\brho\{[h(\bx,\brho)+z]_+\}-z\epsilon\}\le0.
\end{equation}
Each probabilistic constraint in \eqref{eq:P3} will be replaced by the approximation \eqref{eq:approx_max}:
\begin{equation}\label{eq:expValue}
     \mathbb{E}_{\brho^m}\{[h_{i,n}(\bx;\brho^m) + u_{i,n}^m]_+\}-u_{i,n}^m\epsilon \le 0,
\end{equation}
where $u_{i,n}^m$ are real and positive auxiliary optimization variables.
Moreover, since the max operator $[.]_+$ is not differentiable, we replace it with a smooth approximation and define:
\begin{equation}
\begin{aligned}
    &g_{i,n}(\bx,u_{i,n};\brho^m)=\\
    &\quad \frac{1}{2}\left(h_{i,n}(\bx)+u_{i,n}+\sqrt{\xi^2+(h_{i,n}(\bx)+u_{i,n})^2}\right) - u_{i,n}\epsilon \\ 
    &\qquad\forall i \in \{1,2,3,4\}, n \in \cN, m \in \{1,..,b\}   
\end{aligned}
\end{equation}
with $\xi$ small and non-zero. Differentiability of the functions $g_i$ will be required later in Section \ref{sec:convexification} where we will convexify problem (P4) introduced below.
The expected values in \eqref{eq:expValue} can be estimated empirically via sample averaging for a sufficiently large number of samples $N_s$, leading to a new formulation of the optimization problem:
\begin{subequations}\label{eq:P4}
    \begin{align}
        (\text{P4}) \min_{\bx, u_{i,n}^m} &\quad \sum_{m=1}^{b}  \quad  \frac{1}{N_s}\sum_{s=1}^{N_s}h_{0}(\bx;\brho^m[s]) \\ 
        \textrm{s.t.}
        &\quad \frac{1}{N_s}\sum_{s=1}^{N_s} g_{i,n}(\bx, u_{i,n}^m;\brho^m[s]))  \le 0 \nonumber\\
        &\qquad \forall i \in \{1,2,3,4\}, n \in \cN, \, m \in \{1,..,b\} \label{sub:non_convex_set}\\
        &\quad h_{i}(\bx) \leq 0 \quad \forall i \in \{5,6,7,8\} \label{sub:convex_set}
    \end{align}
\end{subequations}
where we will draw $N_s$ samples ${\brho^m[s]}_{s=1}^{N_s}$ of the random vector $\brho^m$. Problem (P4) constitutes a conservative
approximation of the initial chance constrained problem (P3), meaning that an optimal solution to
(P4) is a feasible suboptimal solution to (P3).

\subsection{Solution via successive convex approximation}\label{sec:convexification}

At  first one might try to solve problem (P4) with any software package for nonlinear optimization. However, it is not straightforward to implement the inverse matrix contained in $\bq(\bx)$ in a computationally efficient way. Therefore, we seek a different strategy that may be computationally more affordable. 
In particular, we will leverage the algorithm proposed in~\cite{Scutari2014} which follows the ideas of SCA methods. More specifically, the method solves a sequence of strongly convex inner approximation of an initial non-convex problem. In particular, each intermediate problem is strongly convex and can be written as: 
\begin{equation}
\label{eq:P5}
    \begin{aligned}
        (\text{P5}(\bx_p)) \min_{\bx, u_{i,n}^m} &\quad \sum_{m=1}^{b}  \quad  \frac{1}{N_s}\sum_{s=1}^{N_s}\tilde{h}_{0}(\bx;\brho^m[s],\bx_p) \\
        \textrm{s.t.}
        &\quad \frac{1}{N_s}\sum_{s=1}^{N_s} \tilde{g}_{i,n}(\bx, u_{i,n}^m;\brho^m[s],\bx_p) \le 0\\
        &\qquad \forall i \in \{1,2,3,4\}, n \in \cN, \, m \in \{1,..,b\}\\
        &\quad h_{i}(\bx) \leq 0 \quad \forall i \in \{5,6,7,8\}\\
    \end{aligned}
\end{equation}
where $\tilde{g}_{i,n}(\bx, u_{i,n}^m;\brho^m[s],\bx_p)$ approximates $g_{i,n}(\bx,u_{i,n}^m;\brho^m[s])$ around $\bx=\bx_p$. For given samples $\brho^m[s]$, the problem $(\text{P5}(\bx_p))$ is solved for successive values of $\bx_p$ until convergence. 
The surrogate functions in \eqref{eq:P5} are defined as:
\begin{multline}\label{eq:h0_tilde}
    \tilde{h}_0(\bx;\brho^m[s],\bx_p)=\|\bq(\bx_p)+(\bx-\bx_p)^\top \nabla \bq(\bx_p)\|^2\\
    + \frac{d}{2}\|\bx-\bx_p\|^2
\end{multline}
and 
\begin{equation}\label{eq:g_tilde}
\begin{aligned}
    &\tilde{g}_{i,n}(\bx,u_{i,n}^m;\brho^m[s],\bx_p)=\\
    &\quad \frac{1}{2}\left(\tilde{h}_{i,n}(\bx)+u_{i,n}^m+\sqrt{\xi^2+(\tilde{h}_{i,n}(\bx)+u_{i,n}^m)^2}\right) - u_{i,n}^m\epsilon \\ &\qquad \forall i \in \{1,2,3,4\}, n \in \cN, m \in \{1,..,b\}
\end{aligned}
\end{equation}
with
\begin{equation}\label{eq:hi_tilde}
\begin{aligned}
       &\tilde{h}_{i,n}(\bx;\brho^m[s],\bx_p)=\\
       &h_{i,n}(\bx_p) + (\bx-\bx_p)^\top \nabla h_{i,n}(\bx_p)+(\bx-\bx_p)^\top M_{i,n}(\bx-\bx_p)\\ 
       &\quad \forall i \in \{1,2,3,4\}, n \in \cN, m \in \{1,..,b\} 
\end{aligned}
\end{equation}
where  $M_{i,n} \in \mathbb{R}^{2\times 2}$ is derived according to Appendix~\ref{sec:M_conditions} to ensure that $\tilde{h}_{i,n}(\bx;\brho^m[s],\bx_p)$ is a global majorizer of $h_{i,n}(\bx_p;\brho)$. To lighten the notation, we omit the $\brho^m$ dependence on the right-hand side of equations~(\ref{eq:h0_tilde}--\ref{eq:hi_tilde}) but recall that $\bq(\bx; \brho) = \left[X + \mathbf{1}^\top \bx \mathds{I}\right]^{-1}(\mathds{1}-\brho)$ with $\brho = H(\mathbf{0},\bz)$, and the functions $h_{i,n}(\bx;\brho)$ are defined in equation~\eqref{eq:P2_h_def}. The approximated functions $\tilde{h}_{0}$, $\tilde{g}_{i}$ and $\tilde{h}_{i}$ defined in equations~(\ref{eq:h0_tilde}--\ref{eq:hi_tilde})  must satisfy some assumptions described in~\cite{Scutari2014}, specified in Appendix~\ref{sec: assum_SCA} and verified with our choice of surrogate functions, in order to be suitable for the SCA method.

Next we present our algorithm to solve (P5($\bx_p$)). First, let us define the set $\cK$ defined by equations~\eqref{sub:convex_set}, i.e., the set defined by the convex constraints of problem (P4). Let us also define the set $\cX$ defined by equations~\eqref{sub:convex_set} and~\eqref{sub:non_convex_set}, such that $\cX \subset \cK$. Then, Algorithm~\ref{alg:SCA} is proposed to solve (P5($\bx_p$)).
\begin{algorithm}[h!]
\caption{\emph{Optimal Gain Design (OGD) via SCA}}
\label{alg:SCA}
\textbf{Initialization}: $\gamma_p \in (0,1]$, $\bx_0 \in \cX$. Set $p=0$.\\ 
\quad[1.] Compute the solution $\bx^*(\bx_p)$ of (P5($\bx_p$)).\\
\quad[2.] Set $\bx_{p+1} = \bx_p + \gamma_p(\bx^*(\bx_p)-\bx_p)$\\
\quad[3.] If $\|\bx_{p+1} - \bx_{p}\|< e$ with $e>0$, then STOP.\\
\quad[4.] $p\leftarrow p+1$ and go to step 1.
\end{algorithm}
The convergence properties of Algorithm~\ref{alg:SCA} are based on~\cite[Theorem 1]{Scutari2014}, and are presented in Appendix~\ref{sec: assum_SCA}. 

\begin{remark}{\emph{(Choice of the initial point)}}
Notice that the initial point $\bx_0$ must belong to the set $\cX$. In order to obtain a feasible $\bx_0$, we first start with a small $\eta$ and a large $\alpha$ (0.5 and 3.5, respectively). The equilibrium $\bq^*$ for $\bx = \bx_0$ corresponds to small reactive injections/consumptions such that reactive powers are within the DERs' limits and the stability constraints are met. If the voltage constraints are met, then one can keep $\bx_0$ and does not need to go to step 1 of Algorithm~\ref{alg:SCA}, since it corresponds to a solution with negligible reactive power usage. If the voltage constraints are not met, we slightly increase the value of $\eta$ and decrease the value of $\alpha$, then check the voltage constraints again. We repeat these steps until we find a feasible $\bx_0$. If we cannot find a feasible $\bx_0$, \textit{i.e.}, the set $\cX$ is empty, one should modify the prescribed probability $\epsilon$ to allow for more voltage violations.
\hfill $\Box$
\end{remark}

\section{Implementation of the controllers}\label{sec:implementation_contr}
We assume that each controller is equipped with sensing capabilities, i.e., it is capable of measuring the voltage magnitudes at the node where it is located. For any given DER $g \in \mathcal{G}$ connected to node $n \in \cN$, the following incremental Volt/Var control is implemented:
\begin{subequations}\label{eq:control_implemented}
    \begin{align}
        q_{g,k+1} &= \mathsf{proj}_{\mathcal{Q}_g}[q_{g,k} + \eta(1-v_{n,k}) - (1-\eta)\alpha q_{g,k}]\label{eq:rt_contr_q}\\
        p_{g,k+1} &= \min{\left(p_{g,k},\sqrt{s_{g}^2-q_{g,k+1}^2}\right)},\label{eq:rt_contr_p}
    \end{align}
\end{subequations}
with $s_g$ the nominal rated size of DER $g$.
Equation~\eqref{eq:rt_contr_q} represents the reactive power update of DER $g$ connected at node $n$. The projection operation ensures
that $q_{g,k}$ is always in the feasible set $\mathcal{Q}_g$ of reactive powers for DER $g$.
The approximated voltage $\nu_{n,k}$ written in the initial controller formulation~\eqref{eq:controller} has been replaced by the voltage measurement $v_{n,k}$, which makes this controller fully decentralized (we no longer rely on the matrix $X$, and the impact of other DERs' reactive power is implicitly taken into account through the network feedback). Equation~\eqref{eq:rt_contr_p} indicates that we prioritize reactive power over active power. By prioritizing reactive power over active power, we further mitigate overvoltage issues as the active power injection is reduced and reactive compensation is used. However, this induces active power curtailment which is costly. We will investigate this issue in the next section, where we will present our numerical results. The good behavior of the controller with reactive power prioritization is verified throughout simulations. We leave for future work the theoretical stability study of our controller when considering active power curtailment. 

Note that even if the stability analysis in Section~\ref{sec:feedback_contr_design} does not consider the projection operation of \eqref{eq:control_implemented}, our conclusions in terms of stability are unchanged as long as we consider a single-phase model. Note that the projection map can be decoupled on a per-node basis since $\mathcal{Q} = \mathcal{Q}_1 \times \ldots \times \mathcal{Q}_N$ with $\mathcal{Q}_i$ the set of admissible reactive powers for the $i$th inverter; thus, computed locally at each inverter.
Assuming that we have enough reactive power to regulate voltages, at equilibrium we have that 
\begin{equation}
    \bq^* = \mathsf{proj}_{\mathcal{Q}}[A(\eta,\alpha) \bq^* + B(\eta,\rho)]\, 
\end{equation}
where $A(\eta,\alpha)$ and $B(\eta,\rho)$ were defined in \eqref{eq:controller_la}. This means that $\bq^*$ is a fixed point of the map $\bq \mapsto \mathsf{proj}_{\mathcal{Q}}[A(\eta,\alpha) \bq^* + B(\eta,\rho)]$. Then, we have that 
\begin{equation}
    \begin{aligned}
        \|\bq_{k+1} - \bq^*\| & = \|\mathsf{proj}_{\mathcal{Q}}[A \bq_k + B] - \mathsf{proj}_{\mathcal{Q}}[A \bq^* + B]\| \\
        & \leq \| A \bq_k + B - A\bq^* - B\| \\
        & = \| A \bq_k  - A \bq^* \| \\
        & \leq \| A\| \| \bq_k  - \bq^* \|
    \end{aligned}
\end{equation}
where we used the non-expansiveness property of the projection map in the first inequality, along with standard inequalities of the norms.
We thus obtain: 
\begin{equation}
        \|\bq_{k} - \bq^*\| \leq \| A(\eta,\alpha)\|^k \| \bq_0  - \bq^* \| \, .
\end{equation}
Therefore, the controller is stable if the induced $\ell_2$ norm of the matrix $A(\eta,\alpha)$ satisfies the condition:
\begin{equation}
    \| A(\eta,\alpha)\| < 1 . 
\end{equation}
We recall that if $A$ is symmetric, then $\|A\|= \sqrt{\lambda _{\max }\left(A^{*}A\right)}  = \sqrt{\rho(A)^2} = \rho(A)$.    
Since $X$ is positive definite and symmetric by construction \cite{farivar2013equilibrium}, it follows that $A(\eta,\alpha)$ is symmetric too.
Therefore, we can conclude that if $\rho(A(\eta,\alpha)) < 1$, then the projected controller renders the equilibrium $\bq^*$ globally asymptotically stable. This condition is verified if \eqref{eq:P1_stab_cstr_full} holds, which we have shown in Section \ref{sec:feedback_contr_design}.
\begin{figure}
    \centering
    \includegraphics[width=0.49\textwidth]{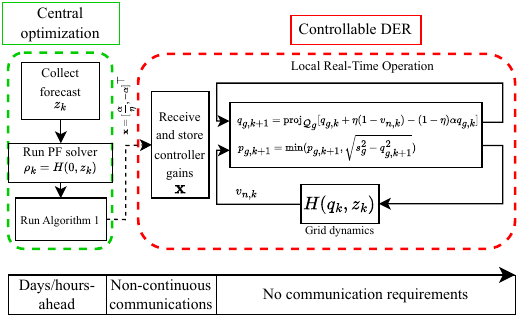}
    \caption{Block diagram of the proposed framework.}
    \label{fig:genframework}
\end{figure}

Fig.~\ref{fig:genframework} illustrates the different stages of the framework proposed in this work.
We compute the gains $\eta$ and $\alpha$ for a given time interval $T\tau$ by solving the problem (P5$(\bx_p)$) until convergence based on forecasts of $\bz_k$. The samples $\brho^m[s]$ are generated by solving multiple power flows for different values of $\bz_k = \bar{\bz}_k + \bdelta_k$, where $\bdelta_k$ follows a given probability distribution function.
The gains can be computed the day before deployment or hours in advance depending on the availability of the forecast and the computational time required to solve (P5$(\bx_p)$) until convergence. They are then broadcast to the controllable DERs. Notice that we do not need to differentiate between DERs as the gains are the same for any DER connected to the network. 

In the following, we address the assumptions introduced in Section~\ref{sec: problem_setup}. First, our current framework enforces one and only one DER per node. This requirement is rather restrictive, even though it has already been adopted in the literature, e.g.~\cite{LiQuDahleh2014}. We can easily relax this assumption by considering only the entries in the matrix $X$ that correspond to the nodes where a controllable DER is located. The drawback is that we can only guarantee voltage satisfaction for a subset $\cN_{\mathrm{red}} \subset \cN$ of nodes. However, the effect of other controllable devices can be embedded in the map $H$. For instance, on-load tap changers (OLTC) or switched capacitor banks can drive the forecast voltage profiles $\brho = H(\mathbf{0},\bz)$ inside the feasible set $\cV$. Our methodology can be combined with other traditional regulation methods, and an optimal combination of slow acting controllers, such as OLTC, with our fast acting controllers is part of our future work.
Furthermore, if multiple DERs are connected to the same node, one can aggregate the DERs and model them as one single device associated with one controller. The reactive power setpoint produced by controller~\eqref{eq:controller} is then appropriately dispatched to the different DERs.

Second, in the present setup the network topology does not change with time. However, in our current framework, topological changes that can be forecast (because of maintenance or planned operation) can easily be integrated. Indeed, those changes impact the matrix $X$. By appropriately choosing the time interval $T\tau$ and recomputing $X$, one can derive gains that would be well adapted to this new network topology. This assumption is much harder to relax for learning-based methods since it requires building new datasets, and learning new equilibrium functions, which can be time-consuming.
When it comes to unplanned changes, e.g. sudden line tripping or unplanned operations, 
our controllers take into account the network conditions as a feedback, and do not worsen the situation. Nevertheless, evaluation of the robustness with respect to unplanned changes remains to be investigated in our future work.

Finally, Assumption~\ref{assumption: main} tackles the feasibility issue of the optimal reactive power flow problem. It may happen that, under our controller architecture, there is not enough reactive power reserve to satisfy the voltage constraints. This problem is implicitly addressed through our chance-constrained formulation. Indeed, increasing the value of $\epsilon$ enlarges the feasible set. For $\epsilon = 1$, the problem (P5$(\bx_p)$) is always feasible.

\section{Numerical experiments}\label{sec:num_exp}

\begin{figure*}[t!]
     \begin{subfigure}[b]{0.31\textwidth}
        \centering
        \includegraphics[width = 1\textwidth]{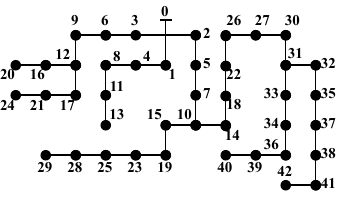}
        \caption{}
        \label{fig:network}
     \end{subfigure}
     \hfill
     \begin{subfigure}[b]{0.31\textwidth}
        \centering
        \includegraphics[width = 1\textwidth]{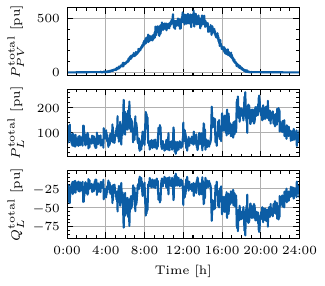}
        \caption{}
        \label{fig:Profiles}
     \end{subfigure}
     \hfill
     \begin{subfigure}[b]{0.31\textwidth}
        \centering
        \includegraphics[width = 1\textwidth]{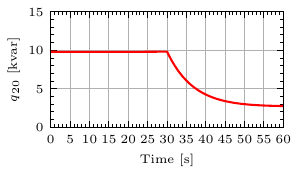}
        \caption{}
        \label{fig:Q_evolv}
     \end{subfigure}
        \caption{(a) Low-voltage 42-nodes network, (b) aggregated non-controllable active power injections, and active/reactive power consumption, (c) reactive power setpoint update for controller at node 20 around hour 20:00.}
\end{figure*}

We consider the low voltage network (0.4 kV) shown in Fig.~\ref{fig:network}. We used a modified network from \cite{SimBench}, in which photovoltaic power (PV) plants have been placed at each node, with inverter-rated size picked randomly among $\{20,25,31\}$ kVA. The DERs dynamics are not implemented, as they are considered to be much faster than the controller dynamics. This is a reasonable assumption because of the time-scale separation between the power system phenomena and the different control loops as mentioned in~\cite{eggli2020stability}. As such, when the controller update law produces a new reactive power setpoint, it is instantaneously implemented by the controllable DER. In this paper, we only consider PV plants as DERs, but any type of inverter-interfaced generation for which the reactive power can be controlled could be considered. Fig.~\ref{fig:Profiles} shows the aggregated loads and maximum available active power for PV plants throughout the day. The data is from the Open Power System 
Data\footnote{Data available at \url{https://data.open-power-system-data.org/household_data/2020-04-15}}, and have been modified to match the initial loads and PV plants nominal values present in the network. The reactive power demand is set such that the power factor is 0.95 (leading). This  represents a typical summer day, with high PV production. We will show that, under these conditions, the electrical DN undergoes both overvoltages and undervoltages.

\subsection{Simulation setup}\label{subsec:sim_setup}
In the following, we assume the controllable DERs to be equipped with the following overvoltage protection.

\emph{Overvoltage protection scheme:} We consider an overvoltage protection for PV plants, i.e., the plant is disconnected from the grid if the voltage goes above 1.06 pu, or stays above 1.05 pu for 10 minutes. The DER reconnects if the voltage remains below 1.05 pu for at least  1 minute. The disconnection scheme is inspired by the CENELEC EN50549-2 standard~\cite{cenelecstandard}, and has been adjusted considering the voltage service limits used in this paper.

In the simulation, the voltage service limits are set to 1.05 and 0.95 p.u., respectively. The load and PV production profiles have a granularity of 1 second, i.e., active/reactive power consumption and maximum available active power for PV plants change every 1 second. The time horizon $T\tau$ is set to 1 hour, and the forecast update $\Delta \tau$ to 30 minutes. The reactive power setpoints update $\tau$ is set to 100 ms.
We compare our proposed controller, based on Algorithm~\ref{alg:SCA} (OGD-CCO), with: 
(a) a static Volt/Var control (VoltVar); and, (b) no control (ON/OFF).

\paragraph{Static Volt/Var control}
It is inspired by the standard \emph{IEEE Std 1547-2018}, with maximum reactive power consumed/absorbed set to 44\% of the nominal power of the DER and reached for voltages 1.05/0.95 pu, respectively. The deadband ranges between 0.99 and 1.01 pu. Active power prioritization is implemented. Therefore, the maximum reactive power that can be produced or consumed corresponds to the minimum between 44\% of the nominal apparent power $s_g$ and the reactive power reserve of the inverter $\sqrt{s^2_g - p^2_{g}}$, with $p_{g}$ the active power injection of DER $g$.

\paragraph{No control} Set controller gains $\eta = \alpha = 0$, and reactive powers to 0.

\subsection{Results} 
\begin{figure*}[h!]
    \begin{subfigure}{.5\textwidth}
        \centering
        \includegraphics{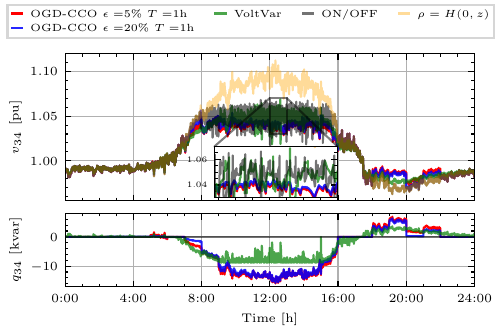}
    \end{subfigure}
    \begin{subfigure}{.5\textwidth}
        \centering
        \includegraphics{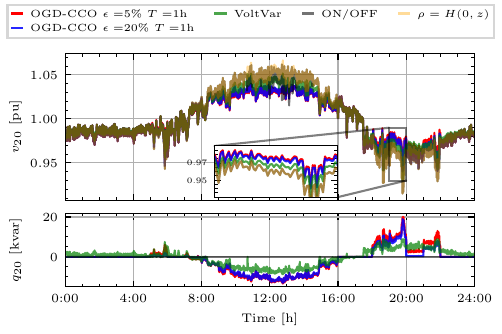}
    \end{subfigure}   
    \caption{Controller dynamics and voltage magnitudes for two given nodes: node 34 and node 20.}
    \label{fig:node}
\end{figure*}

In Fig.~\ref{fig:node}, we illustrate the dynamics of the controllers associated with DER connected to two different nodes and the corresponding voltage magnitudes. 
We compare the voltage magnitudes resulting from the deployment of different controllers against the case without controllers and without overvoltage protection. We fixed $T\tau$ to 1 hour, i.e., the controller gains change every hour. It is clear that the VoltVar control overuses reactive power, as there is no need to control the voltages between hours 16:00 -- 17:00, or after 22:00. The ON/OFF strategy sees large fluctuations in voltages due to constant connection and disconnection of DERs. Fig.~\ref{fig:Q_evolv} illustrates the reactive power setpoint update for the DER connected at node 20 at hour 20:00. In about 7 seconds, the reactive power setpoint varies with a magnitude of 5kvar, which we believe is reasonable. However, if the reactive power setpoints are ramp limited, this would not drastically affect the results since such large variations can occur only during change of hours, and for certain specific changes (in this scenario, at hours 8:00, 16:00, 18:00, 20:00, 21:00 and 22:00).

\begin{figure}[h!]
    \centering
    \includegraphics{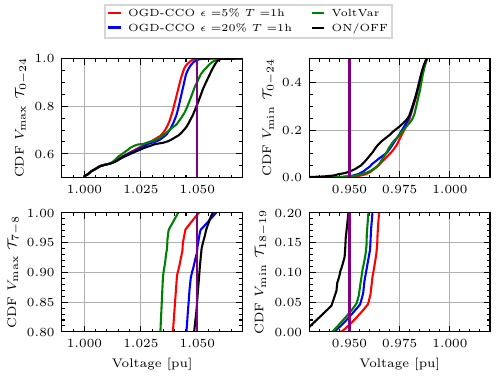}
    \caption{Cumulative distribution function for the vectors $\bV_{\mathrm{max}} = \{\max{v}_{i,k}\}_{i \in \cN, k \in \mathcal{T}_{h1-h2}}$, and $\bV_{\mathrm{min}} = \{\min{v}_{i,k}\}_{i \in \cN, k \in \mathcal{T}_{h1-h2}}$, where $\mathcal{T}_{h1-h2}$ is the set of time indices between hours $h_1$ and $h_2$.}
    \label{fig:CDF}
\end{figure}

The cumulative distribution functions (CDF) for maximum and minimum voltages during the simulation are shown in Fig.~\ref{fig:CDF}. At each time $t$ between hours $h_1$ and $h_2$, we pick the maximum and minimum voltages throughout the network and store them into two distinct vectors. We then plot the cumulative distribution functions of those vectors. For $\cT_{0-24}$, i.e. for the entire duration of the simulation, our proposed method crosses the 1.05 p.u. line above 95\% for the maximum voltage, and lower than 5\% for the undervoltage. This means that the voltage constraints are satisfied at least 95\% of the time, which is consistent with our choice of $\epsilon = 0.05$ and $\epsilon=0.2$. We also show the CDF during two other intervals, between hours 7:00 -- 8:00 for overvoltages and 18:00 -- 19:00 for undervoltages. Notice that a larger $\epsilon$ allows for more voltage constraint violations. The VoltVar control sees overvoltages more than 10\% of the time while performing similarly to our method for undervoltages. The ON/OFF strategy sees overvoltages more than 20\% of the time.

\begin{figure}[h!]
    \centering
    \includegraphics{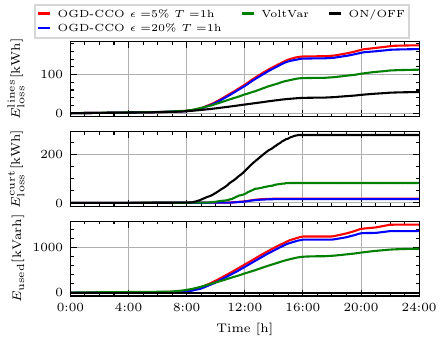}
    \caption{Energy lost in the lines and because of curtailment, and reactive energy usage.}
    \label{fig:e_loss}
\end{figure}

Fig.~\ref{fig:e_loss} displays, from top to bottom, the energy lost in the lines, the energy lost because of active power curtailment, and the cumulative reactive energy usage for the different strategies considered in this work. The line losses are computed based on the line currents squared times the line resistances. Although the usage of reactive power is \emph{practically free}, it induces larger line currents, hence larger power losses. We make a distinction between losses caused by active power curtailment and line losses  since the former are covered by the network user  while the latter are covered by the system operator. However, the system operator can in turn increase network tariffs to compensate for line losses due to overuse of reactive power compensation.

Our controller applies reactive power prioritization, which naturally induces active power curtailment if the DER injects a large amount of active power into the network. On the other hand, the VoltVar and ON/OFF strategies may experience active power curtailment because of prolonged overvoltage violations. Taking all these competing factor into consideration, the total energy loss is much more important for the ON/OFF strategy, while it is equivalent for our controller and the VoltVar control.

These results emphasize the advantages of having gains that are updated to take into account forecast changes in the network conditions. For instance, based on the forecasts, one can observe in Fig.~\ref{fig:node} that the voltages are close to the nominal voltage early in the morning. When computing the gains for that time period, one obtains gains such that the reactive power compensation is almost deactivated as there are no voltage violations forecast. On the other hand, the constant gains of static VoltVar curves are set such that reactive power compensation is provided as soon as the voltage magnitudes deviate and leave the controller deadband, leading to a overusage of reactive power.
One can also see that the parameter $\epsilon$ acts as a lever on the total usage of reactive power. If the reactive power is very expensive, and voltage violations are not extremely important, one could increase the value of $\epsilon$ to reduce the total usage of reactive energy.

\begin{figure}[h!]
    \centering
    \includegraphics{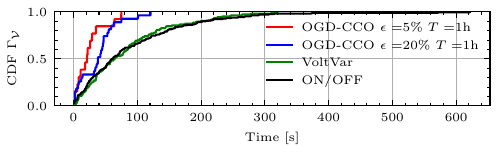}
    \caption{Cumulative distribution function for voltage violations.}
    \label{fig:NumNodes}
\end{figure}

In Fig.~\ref{fig:NumNodes}, we show the maximum duration of voltage violations. The vector $\Gamma_{\cV}$ is built such that, if an overvoltage or undervoltage occurs in the network at a given time step $k\tau$, we count and add up every voltage violations for the subsequent time steps. When, at a subsequent time step, no voltage violation occurs, the total number of voltage violation is appended to the vector, and the counter is reset to 0. The vector $\Gamma_{\cV}$ indicates the duration of voltage violations. This is an important number, since usually electrical devices can cope with short and limited over or under voltages, but may be damaged during prolonged, large excursions from nominal voltages. The parameter $\epsilon$ sets the maximum time the voltage can exceed the voltage limits. 
Fig.~\ref{fig:NumNodes} shows that a smaller $\epsilon$ leads to shorter voltage violations, while both ON/OFF and VoltVar strategies lead to substantially longer voltage violations.

\subsection{Unbalanced three-phase distribution systems}\label{sec:unbalancedDN}
In this section, we show how our algorithm can be naturally applied to a three-phase unbalanced distribution network (the previous sections used a single-phase equivalent in order to make the notation lighter) and present simulation results. For this purpose, we consider the IEEE 123-node test system~\cite{IEEE123}. 

First, we characterize the stability of our control algorithm applied to multi-phase distribution networks. We use the approximation proposed in~\cite{kekatos2015voltage}, where $\bnu_k = X\bq_k + \brho_k$, with $\bnu_k := [(\nu_{1,k}^\phi)^2,...,(\nu_{N,k}^\phi)^2]^\top \in \mathbb{R}^{3N}$, $X \in \mathbb{R}^{3N\times 3N}$, $\bq_k := [q_{1,k}^\phi,...,q_{N,k}^\phi]^\top \in \mathbb{R}^{3N}$, $\brho_k := [(\rho_{1,k}^\phi)^2,...,(\rho_{N,k}^\phi)^2]^\top \in \mathbb{R}^{3N}$, where we use the superscript $\phi$ to denote $(\nu^{\phi}_{n,k})^2 := [(\nu^a_{n,k})^2,(\nu^b_{n,k})^2,(\nu^c_{n,k})^2]$. We follow the approach of~\cite{kekatos2015voltage} for the derivation of the matrix $X$, which is based on the branch flow model and the phase impedance matrices for the lines. Notice that, since it is based on the branch flow model, the approximation considers the squared voltage magnitudes; however, this does not affect our method as the linear approximation (and, thus, the matrix $X$) is used only to design the controller gains and one can simply modify the voltage limits $\bnu_{\text{max}},\bnu_{\text{min}}$ in our optimization-based approach. For this case-study, we assume that DERs are wye-connected, and their reactive powers on each phase can be independently controlled. For the multi-phase case and for a DER $g\in\mathcal{G}$ connected to a node $n\in\mathcal{N}$, the implementation of the controller derived in equations~\eqref{eq:control_implemented} becomes:
\begin{subequations}
    \begin{align}
        q_{g,k+1}^{\phi} &= \text{proj}_{\mathcal{Q}_g^{\phi}}[q^{\phi}_{g,k}+\eta(1-v_{n,k}^\phi)-(1-\eta)\alpha q_{g,k}^\phi]\\
        p_{g,k+1}^\phi &= \text{min}\left(p_{g,k}^\phi,\sqrt{(s^{\phi}_g)^2-(q_{g,k+1}^\phi)^2}\right), 
    \end{align}
\end{subequations}
where $\phi \in \{a,b,c\}$, and where reactive power limits $\mathcal{Q}_g^\phi$ and rated powers $s^\phi_g$ are defined on a per-phase basis. In particular, we consider that our controllable DERs are balanced, share the same rated power $s^\phi_g$ among each phase and thus have the same reactive power limits $\mathcal{Q}_g^\phi$ for each $\phi \in \{a,b,c\}$. Notice that in our simulations, the system is unbalanced because the uncontrollable power injections are not equal among each phase.  

The matrix $X$ of our linear approximation of the three-phase IEEE 123-node test system is positive-definite, which ensures the existence and the uniqueness of the equilibrium. However, it is not symmetric, and, thus, the stability analysis of the projected controller performed in Section~\ref{sec:implementation_contr} no longer holds since $\rho{(A(\eta,\alpha))}\leq \|A(\eta,\alpha)\|$. 
We derive some conditions that highlight the role of $X$; to this end, consider the following inequality:
\begin{equation}
\begin{aligned}
    \|A(\eta,\alpha)\| & = \|(1 - (1 - \eta)\alpha) I - \eta X \| \\
    & \leq \|(1 - (1 - \eta)\alpha) I \| + \|\eta X\| \\
    & \leq |1 - (1 - \eta)\alpha| + \eta \|X\| \, .
\end{aligned}
\end{equation}
Thus, a sufficient condition for the controller to be stable in the multi-phase setup is:
\begin{equation}
     |1 - (1 - \eta)\alpha| + \eta \|X\|< 1 \, 
\end{equation}
which can be written
\begin{equation}
    0 < (1-\eta)\alpha -\eta \|X\|< 2-2\eta \|X\| \,. 
\end{equation}
It yields to the following two constraints when we divide both sides by $\eta$, and introduce the variable $\bx$:
\begin{subequations}
    \begin{align}
        &0 < \one^\top \bx - \|X\|\, \\
        &(\one^\top \bx-2) +\|X\| < 0 \,.
    \end{align}
\end{subequations}
Notice that, as in the single-phase equivalent, we replace $2/\eta$ with 2 such that we obtain a tighter bound on the second constraint as $\eta \in [0,1]$. Although the stability constraints slightly differ from the one presented for the single-phase equivalent, they remain linear with respect to $\mathbf{x}$.

\begin{figure}
    \centering
    \includegraphics[width=.95\linewidth]{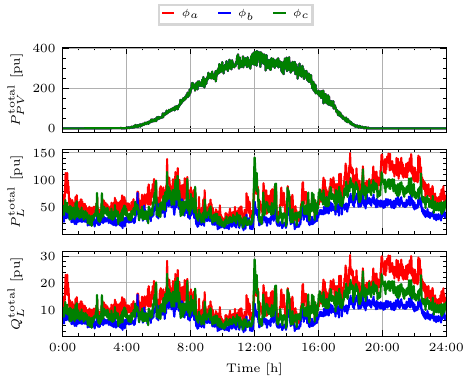}
    \caption{Aggregated non-controllable active power injections, and active/reactive power consumption for the IEEE 123-node test system.}
    \label{fig:Profile_3ph}
\end{figure}

The results obtained on the IEEE 123-node test system are presented next. The simulation setup is the same as the one presented in~\ref{subsec:sim_setup}. The DERs are PV inverters. At each node, a three-phase wye-connected PV inverter is installed, and its rated size is picked randomly among four different inverter sizes $\{60,120,160,200\}$ kVA. The aggregated non-controllable power injections are shown in Fig.~\ref{fig:Profile_3ph}, where one can observe that the system is unbalanced.

\begin{figure}
    \centering
    \includegraphics[width=\linewidth]{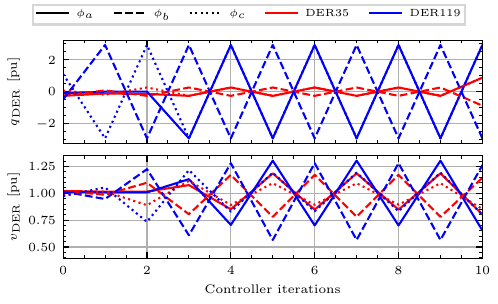}
    \caption{Reactive power updates using the static VoltVar for two DERs in the IEEE 123-node test system.}
    \label{fig:unstable_q}
\end{figure}

Notice that in this multi-phase setup, the static Volt/Var control is unstable and leads to an oscillatory behavior that can be observed in Fig.~\ref{fig:unstable_q}. Therefore, we compare our method for two prescribed probabilities $\epsilon$ with the no control (ON/OFF) strategy. 

\begin{figure}
    \centering
    \includegraphics[width=\linewidth]{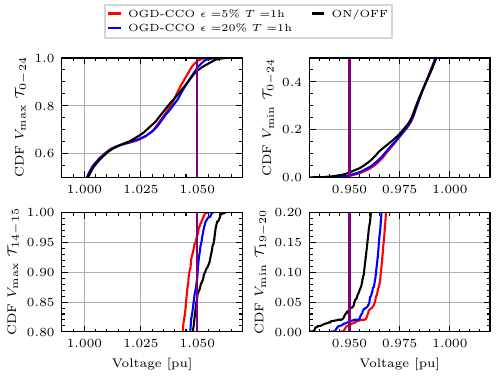}
    \caption{Cumulative distribution function for the vectors $\mathbf{V}_{\text{max}}$ and $\mathbf{V}_{\text{min}}$ for the IEEE 123-node test system.}
    \label{fig:CDF_3ph}
    \vspace{-.3cm}
\end{figure}

\begin{figure}
    \centering
    \includegraphics[width=\linewidth]{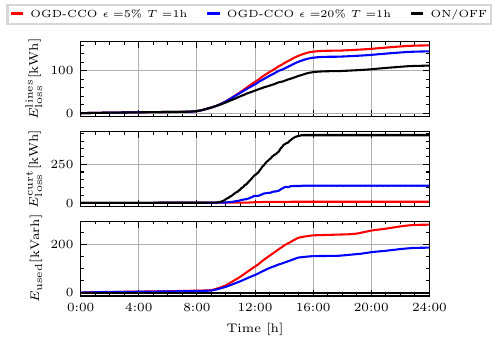}
    \caption{Energy lost in the lines and because of curtailment, and reactive energy usage for the IEEE 123-node test system.}
    \label{fig:power_3ph}
    \vspace{-.1cm}
\end{figure}

The cumulative distribution functions (CDF) for maximum and minimum voltages during the simulation are shown in Fig.~\ref{fig:CDF_3ph}. For the entire duration of the simulation, our method with both prescribed probabilities ($\epsilon = 5\%$ and $\epsilon = 20\%$) performs better than the no control strategy in terms of voltage regulation. Notice that the percentage of voltage violation is consistent with respect to the prescribed probabilities; the CDFs cross the purple line over 95\% and over 90\% for probabilities $\epsilon=5\%$, $\epsilon=20\%$, respectively. 
Fig.~\ref{fig:power_3ph} shows the energy usage of the different methods. One can observe that, while our method uses more reactive energy, it leads to less active power curtailment because of limited voltage violations. One can also see that the total energy lost with the ON/OFF strategy (both coming from active power curtailment and lines losses) is more significant than the energy lost with our method. In particular, our method with a prescribed probability of $\epsilon=5\%$ leads to almost no active power curtailment.
\begin{figure}[t!]
    \centering
    \includegraphics[width=\linewidth]{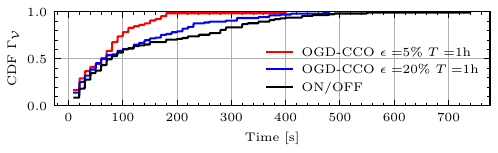}
    \caption{Cumulative distribution function for voltage violations for the IEEE 123-node test system.}
    \label{fig:Numnodebis_3ph}
\end{figure}
Finally, Fig.~\ref{fig:Numnodebis_3ph} shows the maximum duration of voltage violations. When we tolerate less voltage violations (smaller prescribed probability $\epsilon$), the voltage violations are shorter. Notice that in this metric, we consider that there is a voltage violation at time $t$ if at least one voltage among all the nodal voltages for at least one phase is out of the admissible limits. This metric does not indicate the maximum duration of voltage violations for a specific node, but rather provides an overview of the system.

\section{Conclusion}\label{sec:conclusion}
We propose an incremental Volt/Var control strategy for voltage regulation in DNs. We show the stability of our controller and introduce a methodology to compute the gains of our controller based on a chance-constrained formulation of an optimal reactive power flow problem. Our methodology only needs limited and non-continuous communications; that is, the same controller gains are broadcast to individual controllers. Our chance-constrained formulation tackles uncertainties in power injections. Moreover, the feasibility issue of local Volt/Var control, i.e., if there is enough reactive power reserve to satisfy the voltage constraints given the architecture of the controller, is implicitly taken into account by allowing the system operator to tolerate a prescribed probability of voltage violations. Our method shows better performance compared to static Volt/Var curves with fixed parameters, and limited and short voltage violations that are consistent with the prescribed probability.
Our future research will  investigate (i) the combination of our fast-acting controller with slower traditional regulation devices, (ii) the problem of identification of secondary systems and associated uncertainties,  and (iii) considering other network constraints such as maximum line ampacity or maximum allowed voltage unbalance in the design of the gains of the controller.

\appendix 

\subsection{Derivations of $M$}\label{sec:M_conditions}
In this section, we derive the matrices $M_{i,n}$ appearing in~\eqref{eq:hi_tilde} to ensure that $\tilde{h}_{i,n}(\bx;\brho^m[s],\bx_p)$ is a global majorizer of $h_{i,n}(\bx_p;\brho)$. First, let us build the Jacobian matrices $\nabla h_{i}(\bx)$ for $i = 1,2,3,4$ from the Jacobian matrix $\nabla \bq(\bx)$. Notice that the partial derivatives of the vector $\bq(\bx)$ with respect to the first and second component of the vector $\bx$ are identical:
 \begin{equation}\label{eq:derivative}
    \frac{\partial \bq(\bx)}{\partial x_1} = \frac{\partial \bq(\bx)}{\partial x_2} = P \diag{(\gamma(\bx))}P^{-1}(\mathds{1}-\brho)
\end{equation}
where $\gamma(\bx) = [-\frac{1}{(\mathbf{1}^\top \bx+\lambda_{X,1})^2},...,-\frac{1}{(\mathbf{1}^\top \bx+\lambda_{X,N})^2}]^\top$. Since $X$ is positive definite, and $\mathbf{1}^\top \bx \geq 0$ by constraint, the vector $\gamma(\bx)$ has bounded elements.
We can now construct $\nabla \bq(\bx) = \left[\frac{\partial \bq(\bx)}{\partial x_1},\frac{\partial \bq(\bx)}{\partial x_2}\right]^\top \in \mathbb{R}^{2\times N}$ as
\begin{equation}\label{eq:jacobian}
    \begin{aligned}
        \nabla \bq(\bx) = [&P \diag{(\gamma(\bx))}P^{-1}(\mathds{1}-\brho), \\ &\quad P \diag{(\gamma(\bx))}P^{-1}(\mathds{1}-\brho)]^\top.
    \end{aligned}
\end{equation}
Recalling the definitions \eqref{eq:P2_h_def} of the vector-valued functions $h_i$, we can write:
\begin{equation}\label{eq:jacob_h}
    \begin{aligned}
         &\nabla h_1(\bx) =  \nabla \bq(\bx),\quad \nabla h_2(\bx) = - \nabla \bq(\bx)\\
         &\nabla h_3(\bx) = X \nabla \bq(\bx),\quad \nabla h_4(\bx) = -X \nabla \bq(\bx)\\
    \end{aligned}
\end{equation}
In order to determine $M_{i,n}$ such that $M_{i,n}-\nabla^2h_{i,n}(\bx)\succeq	0$, we first derive the Hessian matrix of $h_{i,n}(\bx)$:
\begin{equation}
    \nabla^2h_{i,n}(\bx)= c_{i,n}(\bx) \begin{pmatrix}
1 & 1\\
1 & 1
\end{pmatrix}
\end{equation}
where $c_{i,n}(\bx) \in \mathbb{R}$. 
We define $c_i(\bx) = [c_{i,1}(\bx),...,c_{i,N}(\bx)]^\top$  and from eqs.\eqref{eq:jacobian} and \eqref{eq:jacob_h} it is easy to retrieve:
 \begin{equation}\label{eq:c_h}
    \begin{aligned}
         &c_1(\bx) = P \diag{(d_\gamma(\bx))}P^{-1}(\mathds{1}-\brho), \quad c_2(\bx) = -c_1(\bx)\\
         &c_3(\bx) = X c_1(\bx),\quad c_4(\bx) = -X c_1(\bx)\\
    \end{aligned}
\end{equation}
where $d_\gamma(\bx) = [\frac{2}{(\mathbf{1}^\top \bx+\lambda_{X,1})^3},...,\frac{2}{(\mathbf{1}^\top \bx+\lambda_{X,N})^3}]^\top$.
Given that the quantity $\mathbf{1}^\top \bx$ is always positive, setting $\mathbf{1}^\top\bx=0$ in $d_\gamma(\bx)$ gives us an upper bound on the value of $\nabla^2 h_{i,n}(\bx)$.
Therefore, we write $\bar{c}_i = c_i(\bx)|_{\mathbf{1}^\top \bx = 0}$ for $i=1,2,3,4$ and find:
\begin{equation}
    M_{i,n}=2\begin{pmatrix}
\max(0,\bar{c}_{i,n}) & 0\\
0 & \max(0,\bar{c}_{i,n})
\end{pmatrix}\quad \forall n \in \cN
\end{equation}
such that $\tilde{h}_{i,n}(\bx;\bx_p)$ is a global majorizer of $h_{i,n}(\bx)$ $\forall \bx \in \mathbb{R}^2$.

\subsection{Convergence guarantees for Algorithm~1} \label{sec: assum_SCA}
In this section, we show that the assumptions considered  in~\cite{Scutari2014} for the SCA method to converge are applicable to our problem (P4). To this end, define $\cW \in \mathbb{R}^N$ such that $\brho \in \cW$. 

We start by observing the following properties and facts related to problem (P4). 

\noindent \textbf{[A1]} The set $\cK$ is closed and convex. The set $\cK$ is constructed with linear constraints.

\noindent \textbf{[A2]} The functions $\bx \mapsto h_0(\bx,\brho)$ and $\bx \mapsto g_{i,n}(\bx,u_{i,n};\brho)$ are continuously differentiable on $\cK$, for any $\brho \in \cW$. This follows from the differentiability of $\bq(\bx;\brho)$; since $X$ is positive definite by construction, and $\mathbf{1}^\top \bx\geq 0$ by constraint, it follows that $\nabla \bq(\bx)$ always exists and is bounded. 

\noindent \textbf{[A3]} The gradient $\nabla_{\bx} h_0(\bx;\brho)$ is Lipschitz continuous on $\cK$ with constant $L_{\nabla h_0}$, uniformly in $\brho$. This follows from the fact that $\nabla^2\bq(\bx)$ is bounded on $\cK$.

\noindent \textbf{[A4]} For some $\bx_0 \in \cX$, $\{\bx \in \cX: h_0(\bx;\brho)\leq h_0(\bx_0;\brho)\}$ is compact, for any given $\brho \in \cW$. 

Next, we note that the surrogate function $\tilde{h}_0 : \cK \times \cW \times \cX \rightarrow \mathbb{R}$ is continuously differentiable with respect to $\bx$ and satisfies the following conditions. 

\noindent \textbf{[A5]}  The map $\bx \mapsto \tilde{h}_0(\bx;\brho,\bx_p)$ is  strongly convex on $\cK$ with constant $c\geq d>0$, for any given $\bx_p \in \cX$ and $\brho \in \cW$. 

\noindent \textbf{[A6]} We have that $\nabla \tilde{h}_0(\bx_p;\brho,\bx_p) = \nabla h_0(\bx_p;\brho)$ for any given $\bx_p \in \cX$ and $\brho \in \cW$. to see this, define $f(\cdot) = \|\cdot\|^2$ so and express $h_0(\bx) = f(\bq(\bx))$ and $\tilde{h}_0(\bx;\bx_p) = f(\bq(\bx_p) + (\bx-\bx_p)^\top \nabla \bq(\bx_p)) + \frac{d}{2}\|\bx-\bx_p\|^2$. We can write $\nabla h_0(\bx) = \nabla \bq(\bx)\nabla f(\bq(\bx))$, and $\nabla \tilde{h}_0(\bx;\bx_p) = \nabla \bq(\bx)\nabla f(\bq(\bx) + (\bx-\bx_p)^\top \nabla \bq(\bx)) + d (\bx-\bx_p)$. It is clear that at $\bx=\bx_p$ we have $\nabla \tilde{h}_0(\bx_p;\bx_p)=\nabla h_0(\bx_p)$.

\noindent \textbf{[A7]} The map $(\bx,\bx_p) \mapsto \nabla \tilde{h}_0(\bx;\brho,\bx_p)$ is continuous on $\cK\times \cX$ for any given $\brho \in \cW$.

Finally, the surrogate constraint functions $\tilde{g}_{i,n} : \cK \times \mathbb{R}_{\geq 0} \times \cW \times \cX \rightarrow \mathbb{R}\ \forall i \in \{1,2,3,4\},n\in \cN$ satisfy the following.

\noindent \textbf{[A8]} The map $(\bx,u_{i,n}) \mapsto \tilde{g}_{i,n} (\bx,u_{i,n};\brho,\bx_p)$ is convex on $\cK \times \mathbb{R}_{\geq 0}$ for any given $\brho \in \cW$ and $\bx_p \in \cX$. to see this, note that the composite function $\tilde{g}_{i,n}= f(\tilde{h}_{i,n} + u_{i,n})  - u_{i,n}\epsilon$ is given by the outer function $f(y)=\frac{1}{2}\left(y+\sqrt{\xi^2+(y)^2}\right)$ which is strictly increasing and convex and by the inner function $\tilde{h}_{i,n} + u_{i,n}$ which is convex by construction; it then follows that $\tilde{g}_{i,n}$ is also convex.

\noindent \textbf{[A9]} $\tilde{g}_{i,n} (\bx_p,u_{i,n};\brho,\bx_p) = g_{i,n} (\bx_p,u_{i,n};\brho)$ for any given $\bx_p \in \cX$, $u_{i,n} \in \mathbb{R}_{\geq 0}$ and $\brho \in \cW$.

\noindent \textbf{[A10]} $g_{i,n} (\bx,u_{i,n};\brho) \leq \tilde{g}_{i,n} (\bx,u_{i,n};\brho,\bx_p)$ for all $\bx \in \cK$, $u_{i,n} \in \mathbb{R}_{\geq 0}$ and for any given $\bx_p \in \cX$, $\brho \in \cW$.

\noindent \textbf{[A11]} The map $(\bx,u_{i,n},\bx_p) \mapsto \tilde{g}_{i,n} (\bx,u_{i,n};\brho,\bx_p)$ is continuous on $\cK \times \mathbb{R}_{\geq 0} \times \cX$ for any given $\brho \in \cW$.

\noindent \textbf{[A12]} $ \nabla \tilde{g}_{i,n}(\bx_p,u_{i,n,p}^{m}; \brho, \bx_p) = \nabla g_{i,n}(\bx_p,u_{i,n,p}^{m}; \brho)$ for all $\bx_p \in \cX$, $ u_{i,n,p}^{m} \in \mathbb{R}_{\geq 0}$ and for any given $\bx_p \in \cX$, $\brho \in \cW$. To see this, using the chain rule we obtain
\begin{equation}
\begin{aligned}
    \frac{\partial g_{i,n}(\bx,u)}{\partial \bx} &= \frac{1}{2}\left(1+\frac{h_{i,n}(\bx) + u}{\sqrt{\xi^2+(h_{i,n}(\bx)+u)^2}}\right)\frac{\partial h_{i,n}}{\partial \bx},\\
\frac{\partial g_{i,n}(\bx,u)}{\partial u} &= \frac{1}{2}\left(1+\frac{h_{i,n}(\bx) + u}{\sqrt{\xi^2+(h_{i,n}(\bx)+u)^2}}\right)- \epsilon.
\end{aligned}
\end{equation}
The claim is verified since $\frac{\partial \tilde{h}_{i,n}}{\partial \bx}(\bx_p;\brho,\bx_p) = \frac{\partial h_{i,n}}{\partial \bx}(\bx_p;\brho)$ and $\tilde{h}_{i,n}(\bx_p;\brho,\bx_p) = h_{i,n}(\bx_p;\brho)$ for all $\bx_p \in \cX$ and any given $\brho \in \cW$.

\noindent \textbf{[A13]}  The map $(\bx,u_{i,n},\bx_p) \mapsto \nabla \tilde{g}_{i,n} (\bx,u_{i,n};\brho,\bx_p)$ is continuous on $\cK \times \mathbb{R}_{\geq 0} \times \cX$ for any given $\brho \in \cW$.

Given the above properties [A1]--[A13], the following holds when using Algorithm~\ref{alg:SCA} to solve (P4).

\begin{proposition} 
\label{prop:convergence}
Consider the problem (P4), and use Algorithm~\ref{alg:SCA} with a step-size $\gamma_p$. Assume that one of the following two conditions is satisfied:
\begin{itemize}
    \item $0 < \inf_{p} \gamma_p \leq \sup_{p} \gamma_p \leq \gamma^{\mathrm{max}} \leq 1$ and $2c\geq \gamma^{\mathrm{max}}L_{\nabla h_0}$,
    \item $\cX$ is compact, $\bx^*(\bx_p) \in \cX$ is regular for every $\bx_p \in \cX$ and $\gamma^p \in (0,1],\ \gamma_p \rightarrow 0,\ \sum_{p}\gamma_p = +\infty$.
\end{itemize}
Then, every regular limit point of $\{\bx_p\}$ is a stationary solution of (P4). Furthermore, none of such points is a local maximum of $h_0(\bx)$.  \hfill $\Box$
\end{proposition}

The result of Proposition~\ref{prop:convergence} follows from~\cite[Theorem 1]{Scutari2014}. It asserts that the Algorithm~\ref{alg:SCA} converges to a stationary solution of (P4), and this solution is guaranteed not to be a local maximum. 

\bibliographystyle{IEEEtran}
\bibliography{references}

\begin{thebibliography}{10}
\providecommand{\url}[1]{#1}
\csname url@samestyle\endcsname
\providecommand{\newblock}{\relax}
\providecommand{\bibinfo}[2]{#2}
\providecommand{\BIBentrySTDinterwordspacing}{\spaceskip=0pt\relax}
\providecommand{\BIBentryALTinterwordstretchfactor}{4}
\providecommand{\BIBentryALTinterwordspacing}{\spaceskip=\fontdimen2\font plus
\BIBentryALTinterwordstretchfactor\fontdimen3\font minus \fontdimen4\font\relax}
\providecommand{\BIBforeignlanguage}[2]{{%
\expandafter\ifx\csname l@#1\endcsname\relax
\typeout{** WARNING: IEEEtran.bst: No hyphenation pattern has been}%
\typeout{** loaded for the language `#1'. Using the pattern for}%
\typeout{** the default language instead.}%
\else
\language=\csname l@#1\endcsname
\fi
#2}}
\providecommand{\BIBdecl}{\relax}
\BIBdecl

\bibitem{walling2008summary}
R.~Walling, R.~Saint, R.~C. Dugan, J.~Burke, and L.~A. Kojovic, ``Summary of distributed resources impact on power delivery systems,'' \emph{IEEE Trans. on Power Delivery}, vol.~23, no.~3, pp. 1636--1644, 2008.

\bibitem{alam2020high}
M.~S. Alam, F.~S. Al-Ismail, A.~Salem, and M.~A. Abido, ``High-level penetration of renewable energy sources into grid utility: Challenges and solutions,'' \emph{IEEE Access}, vol.~8, pp. 190\,277--190\,299, 2020.

\bibitem{mcdonald2013solar}
J.~McDonald, ``Solar power impacts power electronics in the smart grid,'' \emph{Power Electronics}, 2013.

\bibitem{antoniadou2017distributed}
K.~E. Antoniadou-Plytaria, I.~N. Kouveliotis-Lysikatos, P.~S. Georgilakis, and N.~D. Hatziargyriou, ``Distributed and decentralized voltage control of smart distribution networks: Models, methods, and future research,'' \emph{IEEE Trans. on Smart Grid}, vol.~8, no.~6, pp. 2999--3008, 2017.

\bibitem{Zhu2016}
H.~Zhu and H.~J. Liu, ``Fast local voltage control under limited reactive power: Optimality and stability analysis,'' \emph{IEEE Trans. on Power Syst.}, vol.~31, no.~5, pp. 3794--3803, 2016.

\bibitem{Zhou2021}
X.~Zhou, M.~Farivar, Z.~Liu, L.~Chen, and S.~H. Low, ``Reverse and forward engineering of local voltage control in distribution networks,'' \emph{IEEE Trans. on Automatic Control}, vol.~66, no.~3, pp. 1116--1128, 2021.

\bibitem{Jahangiri2013}
P.~Jahangiri and D.~C. Aliprantis, ``Distributed {V}olt/{VAr} control by {PV} inverters,'' \emph{IEEE Trans. on Power Syst.}, vol.~28, no.~3, pp. 3429--3439, 2013.

\bibitem{Farivar2015}
M.~Farivar, X.~Zho, and L.~Chen, ``Local voltage control in distribution systems: An incremental control algorithm,'' in \emph{2015 IEEE International Conference on Smart Grid Communications (SmartGridComm)}, 2015, pp. 732--737.

\bibitem{Bolognani2019}
S.~Bolognani, R.~Carli, G.~Cavraro, and S.~Zampieri, ``On the need for communication for voltage regulation of power distribution grids,'' \emph{IEEE Trans. on Control of Network Systems}, vol.~6, no.~3, pp. 1111--1123, 2019.

\bibitem{Baker2017}
K.~Baker, A.~Bernstein, E.~Dall’Anese, and C.~Zhao, ``Network-cognizant voltage droop control for distribution grids,'' \emph{IEEE Trans. on Power Syst.}, vol.~33, no.~2, pp. 2098--2108, 2018.

\bibitem{Wei2023}
J.~Wei, S.~Gupta, D.~C. Aliprantis, and V.~Kekatos, ``A chance-constrained optimal design of {V}olt/{VAR} control rules for distributed energy resources,'' in \emph{2023 North American Power Symposium (NAPS)}, 2023, pp. 1--6.

\bibitem{nazir2019}
F.~U. Nazir, B.~C. Pal, and R.~A. Jabr, ``A two-stage chance constrained volt/var control scheme for active distribution networks with nodal power uncertainties,'' \emph{IEEE Trans. on Power Syst.}, vol.~34, no.~1, pp. 314--325, 2019.

\bibitem{Karagiannopoulos2019}
S.~Karagiannopoulos, P.~Aristidou, and G.~Hug, ``Data-driven local control design for active distribution grids using off-line optimal power flow and machine learning techniques,'' \emph{IEEE Trans. on Smart Grid}, vol.~10, no.~6, pp. 6461--6471, 2019.

\bibitem{eggli2020stability}
A.~Eggli, S.~Karagiannopoulos, S.~Bolognani, and G.~Hug, ``Stability analysis and design of local control schemes in active distribution grids,'' \emph{IEEE Trans. on Power Syst.}, vol.~36, no.~3, pp. 1900--1909, 2020.

\bibitem{Gupta2023}
S.~Gupta, A.~Mehrizi-Sani, S.~Chatzivasileiadis, and V.~Kekatos, ``Deep learning for scalable optimal design of incremental {V}olt/{VAR} control rules,'' \emph{IEEE Control Systems Letters}, vol.~7, pp. 1957--1962, 2023.

\bibitem{gupta2024optimal}
S.~Gupta, V.~Kekatos, and S.~Chatzivasileiadis, ``Optimal design of {V}olt/{VAR} control rules of inverters using deep learning,'' \emph{arXiv:2211.09557}, 2024.

\bibitem{yuan2023learning}
Z.~Yuan, G.~Cavraro, M.~K. Singh, and J.~Cortés, ``Learning provably stable local {V}olt/{Var} controllers for efficient network operation,'' \emph{IEEE Trans. on Power Syst.}, vol.~39, no.~1, pp. 2066--2079, 2024.

\bibitem{SERNATORRE2022108575}
P.~{Serna Torre} and P.~Hidalgo-Gonzalez, ``Decentralized optimal power flow for time-varying network topologies using machine learning,'' \emph{Electric Power Systems Research}, vol. 212, p. 108575, 2022.

\bibitem{Nemirovski2007}
A.~Nemirovski and A.~Shapiro, ``Convex approximations of chance constrained programs,'' \emph{SIAM Journal on Optimization}, vol.~17, no.~4, pp. 969--996, 2007.

\bibitem{Scutari2014}
G.~Scutari, F.~Facchinei, L.~Lampariello, and P.~Song, ``Parallel and distributed methods for nonconvex optimization,'' in \emph{2014 IEEE International Conference on Acoustics, Speech and Signal Processing (ICASSP)}, 2014, pp. 840--844.

\bibitem{bolognani2015existence}
S.~Bolognani and S.~Zampieri, ``On the existence and linear approximation of the power flow solution in power distribution networks,'' \emph{IEEE Trans. on Power Syst.}, vol.~31, no.~1, pp. 163--172, 2015.

\bibitem{wang2016explicit}
C.~Wang, A.~Bernstein, J.-Y. Le~Boudec, and M.~Paolone, ``Explicit conditions on existence and uniqueness of load-flow solutions in distribution networks,'' \emph{IEEE Trans. on Smart Grid}, vol.~9, no.~2, pp. 953--962, 2016.

\bibitem{wang2017existence}
------, ``Existence and uniqueness of load-flow solutions in three-phase distribution networks,'' \emph{IEEE Trans. on Power Syst.}, vol.~32, no.~4, pp. 3319--3320, 2017.

\bibitem{dall2016optimal}
E.~Dall’Anese and A.~Simonetto, ``Optimal power flow pursuit,'' \emph{IEEE Trans. on Smart Grid}, vol.~9, no.~2, pp. 942--952, 2016.

\bibitem{LiQuDahleh2014}
N.~Li, G.~Qu, and M.~Dahleh, ``Real-time decentralized voltage control in distribution networks,'' in \emph{2014 52nd Annual Allerton Conference on Communication, Control, and Computing (Allerton)}, 2014, pp. 582--588.

\bibitem{baran1989network}
M.~E. Baran and F.~F. Wu, ``Network reconfiguration in distribution systems for loss reduction and load balancing,'' \emph{IEEE Trans. on Power delivery}, vol.~4, no.~2, pp. 1401--1407, 1989.

\bibitem{farivar2013equilibrium}
M.~Farivar, L.~Chen, and S.~Low, ``Equilibrium and dynamics of local voltage control in distribution systems,'' in \emph{52nd IEEE Conference on Decision and Control}, 2013, pp. 4329--4334.

\bibitem{kekatos2015voltage}
V.~Kekatos, L.~Zhang, G.~B. Giannakis, and R.~Baldick, ``Voltage regulation algorithms for multiphase power distribution grids,'' \emph{IEEE Trans. on Power Syst.}, vol.~31, no.~5, pp. 3913--3923, 2015.

\bibitem{SimBench}
D.~Sarajlić and C.~Rehtanz, ``Low voltage benchmark distribution network models based on publicly available data,'' in \emph{2019 IEEE PES Innovative Smart Grid Technologies Europe (ISGT-Europe)}, 2019, pp. 1--5.

\bibitem{cenelecstandard}
``Requirements for generating plants to be connected in parallel with distribution networks – {P}art 2: connection to a {MV} distribution network – generating plants up to and including type {B},'' \emph{CENELEC}, pp. 1--84, 2019.

\bibitem{IEEE123}
\BIBentryALTinterwordspacing
IEEE. (1992) {IEEE PES} test feeders. [Online]. Available: \url{https://cmte.ieee.org/pes-testfeeders/resources/}
\BIBentrySTDinterwordspacing

\end{thebibliography}

\end{document}